\documentclass[12pt]{article}
\usepackage{amsmath,amssymb,amsfonts,amsthm,bm,bbm,cancel,wasysym}
\usepackage{epsfig,graphics,graphicx,epstopdf,caption,subcaption}
\usepackage[numbers,sort&compress]{natbib}
\graphicspath{{Charts/}}
\usepackage{array,booktabs,colortbl,colordvi,multirow}
\usepackage{colordvi,color,xcolor}
\usepackage{hyperref}
\usepackage{rotating}
\usepackage{comment}
\usepackage{feynmf}

\usepackage[symbol]{footmisc}

\RequirePackage[normalem]{ulem} 
\RequirePackage{color}\definecolor{RED}{rgb}{1,0,0}\definecolor{BLUE}{rgb}{0,0,1} 
\providecommand{\DIFaddtex}[1]{{\protect\color{red}#1}} 
\providecommand{\DIFaddbegin}{\begingroup\color{red}} 
\providecommand{\DIFaddend}{\endgroup} 
\providecommand{\DIFadd}[1]{\texorpdfstring{\DIFaddtex{#1}}{#1}} 
\makeatletter 
\let\sout@orig\sout 
\renewcommand{\sout}[1]{\ifmmode\text{\sout@orig{\ensuremath{#1}}}\else\sout@orig{#1}\fi} 
\makeatother 
\RequirePackage{listings} 
\RequirePackage{color} 
\lstdefinelanguage{DIFcode}{ 
  moredelim=[il][\color{red}\sout]{\%DIF\ <\ }, 
  moredelim=[il][\color{red}\uwave]{\%DIF\ >\ } 
} 
\lstdefinestyle{DIFverbatimstyle}{ 
	language=DIFcode, 
	basicstyle=\ttfamily, 
	columns=fullflexible, 
	keepspaces=true 
} 
\lstnewenvironment{DIFverbatim}{\lstset{style=DIFverbatimstyle}}{} 
\lstnewenvironment{DIFverbatim*}{\lstset{style=DIFverbatimstyle,showspaces=true}}{} 
\begin{document}

\begin{center}

{\Large\bfseries 21-cm Constraints on S-wave Dark Matter Annihilation with Subhalo-enhanced Effects}

\vspace*{0.75cm}

{Zixuan Xu$^{a,}$\footnote{zixuanxu2010@gmail.com} and Sibo Zheng$^{b,}$\footnote{sibozheng.zju@gmail.com}}

\vspace{0.5cm}
{$^{a}$Department of Mathematics and Physics, Luoyang Institute of Science and Technology, Luoyang, Henan 471023, China \\
$^{b}$School of Physics, Chongqing University, Chongqing 401331, China}
\end{center}
\vspace{.5cm}


\begin{abstract}
\noindent
Observations of the 21-cm signal can probe energy injection into the intergalactic medium arising from dark matter annihilation during cosmic dawn and reionization.
For $s$-wave annihilation, the annihilation rate depends on the squared dark matter density,
making the 21-cm signal sensitive to dark matter halo substructure.
In this work, we extend the numerical method of \texttt{DM21cm} to address spatially inhomogeneous $s$-wave annihilation including subhalo effects, 
using semi-analytical treatment of SASHIMI-C.
Our results show that the projected HERA limits for the $\gamma\gamma$ ($e^+e^-$) annihilation channel are the strongest in the dark matter mass range of $m_\chi\sim0.1$--$3\,{\rm keV}$ ($m_\chi\sim 1$ MeV --$3$ GeV) compared to existing bounds.
\end{abstract}

\thispagestyle{empty}
\vfill
\newpage
\setcounter{page}{1}

\tableofcontents

\section{Introduction}
\label{sec:introduction}

The hyperfine transition of neutral hydrogen in the intergalactic medium (IGM) offers a powerful means of tracing the evolution of the Universe from the cosmic dark ages to the epoch of reionization~\cite{Furlanetto:2006jb,Pritchard:2011xb}. 
The neutral hydrogen fraction and spin temperature directly affect fluctuations in the 21-cm brightness-temperature field and power spectra
that can be probed by radio interferometers such as the Hydrogen Epoch of Reionization Array (HERA) \cite{HERA:2021bsv,HERA:2022wmy}
and the Square Kilometer Array (SKA) \cite{Koopmans:2015sua}. 
On the other hand, the ionization and thermal state of the IGM are shaped by cosmological and astrophysical sources of heating, ionization, and excitation.
In this sense, the 21-cm cosmology provides a window to identify those sources in the late-time Universe.

Apart from the above sources,
dark matter (DM) may also affect the thermal and ionization state of the IGM.
Energetic photons and electrons produced by DM annihilation or decay can deposit
additional energy into the gas, modifying its temperature, ionization fraction, and Lyman-$\alpha$ radiation field.
Therefore, the 21-cm observations can be used to constrain DM annihilation and decay \cite{DAmico:2018sxd,Xu:2024uas,Sun:2023acy,Cima:2025zmc,Natwariya:2025jlw,Sun:2025ksr,Cang:2023bnl}.
Most of these recent studies rely on the numerical framework of \texttt{DM21cm}~\cite{Sun:2023acy}, 
which combines \texttt{DarkHistory}~\cite{Liu:2019bbm} and \texttt{21cmFAST}~\cite{Mesinger:2010ne,Qin:2020xyh} to model spatially inhomogeneous energy injection and its impact on the 21-cm signal.
In particular, Ref.~\cite{Natwariya:2025jlw} studied $s$-wave DM annihilation into $e^+e^-$ and $\gamma\gamma$, 
but analyzed the DM density field on the cosmological simulation grid, where subhalo effects were neglected.
Ref.~\cite{Sun:2025ksr} developed a halo-based treatment for $p$-wave DM annihilation, in which subhalo effects are small due to velocity suppression.

Unlike in the $p$-wave case, subhalo effects are not suppressed by velocity for $s$-wave DM annihilation.
The contribution from the subhalo population should therefore be taken into account~\cite{Hiroshima:2018kfv,Ando:2019xlm,Hiroshima:2022khy}.
This is the main task of this work. 
We will extend the work of \texttt{DM21cm} to include the subhalo effects in the case of $s$-wave DM annihilation.
To do so, we first model the subhalo population, using the Semi-Analytical SubHalo Inference ModelIng for Cold dark matter (SASHIMI-C) framework \cite{Hiroshima:2018kfv,Ando:2019xlm}, 
to account for the enhancement on host-halo annihilation luminosity, which depends on the host halo mass and redshift.
Then, we discuss how subhalo structure affects spatially inhomogeneous $s$-wave DM annihilation through the annihilation luminosity.
Finally, we implement these effects in \texttt{DM21cm} to study their impacts on the 21-cm observables.

The remainder of this paper is organized as follows.
In Sec.~\ref{sec:swave_halo}, we present the halo-based treatment of
$s$-wave DM annihilation without and with the subhalos.
In Sec.~\ref{sec:impact}, we study the impacts of s-wave DM annihilation on the thermal and ionization histories of IGM and the 21-cm observables,
with emphasis on the subhalo effects.
In Sec.~\ref{sec:constraints}, we present the projected HERA constraints, which are compared to existing bounds.
Finally, we conclude in Sec.~\ref{sec:conclusion}.

Note: throughout this paper we use the explicit values of cosmological parameters $\Omega_m=0.30966$, $\Omega_b=0.04897$, $h=0.6766$, $n_s=0.9665$, and $\sigma_8=0.8102$ \cite{Planck:2018vyg},
where $\Omega_m$ is the matter density parameter, 
$\Omega_b$ is the baryon density parameter,  
$h=H_0/(100\,{\rm km\,s^{-1}\,Mpc^{-1}})$, 
$n_s$ is the primordial scalar spectral index, 
and $\sigma_8$ is the linear matter fluctuation amplitude on $8h^{-1}\,{\rm Mpc}$ scales.

\section{S-wave DM annihilation in halos}
\label{sec:swave_halo}

\subsection{Annihilation luminosity in smooth DM halos}
\label{subsec:smooth_halo}
For self-conjugate DM, the annihilation rate in a halo is
\begin{equation}
\Gamma_{\rm host}=\frac{1}{2}\int dV\,\langle \sigma v\rangle n_{\rm DM}^2 ,
\label{eq:gamma_general}
\end{equation}
where $n_{\rm DM}$ is the DM number density, $\langle \sigma v\rangle$ is the velocity-averaged DM annihilation cross section,
and the factor $1/2$ arises from the assumption that the DM particle is self-conjugate such as a Majorana fermion.
For nonrelativistic DM, the annihilation cross section can be expanded in powers of the relative velocity as
\begin{equation}
\langle \sigma v\rangle=a+b\frac{\langle v_{\rm rel}^2\rangle}{c^2}+\mathcal{O}\left(\frac{v_{\rm rel}^4}{c^4}\right),
\label{eq:partial_wave}
\end{equation}
where the first and second terms correspond to the leading $s$-wave and $p$-wave contributions, respectively.
In this work, we consider $s$-wave annihilation only,
which is independent of the DM velocity.

To determine the annihilation rate in Eq.(\ref{eq:gamma_general}),  
we model the smooth matter density of the halo with a Navarro--Frenk--White (NFW) profile \cite{Navarro:1995iw},
\begin{equation}
\rho_{\rm NFW}(r)=\frac{\rho_s}{(r/r_s)(1+r/r_s)^2},
\label{eq:nfw}
\end{equation}
where $\rho_s$ and $r_s$ are the characteristic density and scale radius, respectively.
We define the halo mass $M_{200}$ as the mass enclosed within $r_{200}$,
where the mean enclosed density is 200 times the critical density at redshift $z$,
and define the concentration as $c_{200}=r_{200}/r_s$ \cite{Mo:2010ga}.
We use the concentration--mass relation $c_{200}(M,z)$ from Ref.\cite{Ludlow:2016ifl}, 
as implemented in \texttt{halomod} \cite{Murray:2013qza,Murray:2020dcd}.

Using $n_{\rm DM}=[\Omega_{\rm DM}/(\Omega_{\rm DM}+\Omega_b)]\rho_{\rm NFW}/m_\chi$ with $m_\chi$ the DM mass, 
$\Omega_{\rm DM}$ and $\Omega_b$ the present-day DM and baryon density parameters respectively, and $\rho_{\rm NFW}$ given by Eq.(\ref{eq:nfw}), 
Eq.(\ref{eq:gamma_general}) can be rewritten as \cite{Sun:2025ksr}
\begin{equation}
\Gamma_{\rm host}^{(0)}(M,z)=\frac{\langle\sigma v\rangle}{2m_\chi^2}
\left(\frac{\Omega_{\rm DM}}{\Omega_{\rm DM}+\Omega_b}\right)^2
\int_0^{r_{200}} 4\pi r^2dr\,\rho_{\rm NFW}^2(r|M,z),
\label{eq:gamma_swave}
\end{equation} 
In contrast to the $p$-wave case \cite{Sun:2025ksr},
the annihilation rate in Eq.(\ref{eq:gamma_swave}) is not suppressed by velocity in the $s$-wave case.

Given the annihilation rate in Eq.(\ref{eq:gamma_swave}), 
the annihilation luminosity reads as 
\begin{equation}
L_{\rm host}(M,z)=2m_\chi\Gamma_{\rm host}^{(0)}(M,z).
\label{eq:Lhost}
\end{equation}

\subsection{Subhalo-enhanced effects on the annihilation luminosity}
\label{subsec:subhalo_luminosity}

Host halos contain populations of gravitationally bound subhalos 
over a broad range of masses~\cite{Springel:2008cc,Diemand:2008in,Ando:2019xlm}.
Since the $s$-wave annihilation rate is proportional to $\rho_\chi^2$,
the presence of dense substructures enhances the annihilation luminosity relative to 
that of a completely smooth host halo~\cite{Ando:2019xlm}.

Following \cite{Ando:2019xlm}, we decompose the DM density within a host halo as
\begin{equation}
\rho_\chi(\mathbf r)=\rho_{\rm sm}(\mathbf r)+\rho_{\rm sh}(\mathbf r),
\label{eq:rho_sub_decomp}
\end{equation}
where $\rho_{\rm sm}$ denotes the smooth host component  and $\rho_{\rm sh}$ is the density of subhalos.
This decomposition gives
\begin{equation}
\int dV\,\rho_\chi^2=\int dV\,
\left(\rho_{\rm sm}^2+\rho_{\rm sh}^2+2\rho_{\rm sm}\rho_{\rm sh}\right).
\label{eq:rho2_sub_decomp}
\end{equation}

Furthermore, we define the subhalo fraction of the host halo mass as~\cite{Ando:2019xlm}
\begin{equation}
f_{\rm sh}(M,z)=\frac{1}{M}\int dm\,m\,\frac{dN_{\rm sh}(m|M,z)}{dm},
\label{eq:fsh}
\end{equation}
where $M$ is the host halo mass, $m$ is the subhalo mass, and
$dN_{\rm sh}(m|M,z)/dm$ is the subhalo mass function, i.e. the
differential number of subhalos of mass $m$. 
Using the mass fraction in Eq.(\ref{eq:fsh}), 
the smooth and mean subhalo densities in Eq.(\ref{eq:rho_sub_decomp}) can be written as $\rho_{\rm sm}=(1-f_{\rm sh})[\Omega_{\rm DM}/(\Omega_{\rm DM}+\Omega_b)]\rho_{\rm NFW}$ and $\langle\rho_{\rm sh}\rangle=f_{\rm sh}[\Omega_{\rm DM}/(\Omega_{\rm DM}+\Omega_b)]\rho_{\rm NFW}$, respectively. 

We can express the contribution to the annihilation luminosity due to the subhalo population in terms of the subhalo boost factor~\cite{Ando:2019xlm},
\begin{equation}
L_{\rm sub}=B_{\rm sh}L_{\rm host}(M,z).
\label{eq:Lsub_subhalo}
\end{equation}
with 
\begin{equation}
B_{\rm sh}(M,z)=\frac{1}{L_{\rm host}(M,z)}\int dm\,\frac{dN_{\rm sh}(m|M,z)}{dm}\,L_{\rm sh}(m),
\label{eq:Bsh}
\end{equation}
where $L_{\rm sh}(m)$ denotes the annihilation luminosity of an individual subhalo of mass $m$ residing in the host halo.

Replacing the integral in Eq.(\ref{eq:gamma_swave}) with Eq.(\ref{eq:rho2_sub_decomp}), 
the annihilation luminosity with subhalo effects taken into account is now given by \cite{Ando:2019xlm}
\begin{equation}
L_{\rm total}(M,z)=L_{\rm sm}+L_{\rm sub}+L_{\rm cross}
=\left[1-f_{\rm sh}^2(M,z)+B_{\rm sh}(M,z)\right]L_{\rm host}(M,z).
\label{eq:Lhost_subhalo}
\end{equation}
where Eq.(\ref{eq:Lsub_subhalo}) has been used, and the annihilation luminosities  
\begin{equation}
L_{\rm sm}=(1-f_{\rm sh})^2 L_{\rm host}(M,z),~~
L_{\rm cross}=2f_{\rm sh}(1-f_{\rm sh})L_{\rm host}(M,z).
\end{equation}
Dividing Eq.~(\ref{eq:Lhost_subhalo}) by Eq.~(\ref{eq:Lhost}), one obtains the subhalo enhancement factor for the annihilation luminosity,
\begin{equation}
B(M,z)
\equiv\frac{L_{\rm total}(M,z)}{L_{\rm host}(M,z)}=1-f_{\rm sh}^2(M,z)+B_{\rm sh}(M,z),
\label{eq:Btotal}
\end{equation}
with $B=1$ corresponding to the smooth host limit.

Evaluating the $B$ factor in Eq.~(\ref{eq:Btotal}) requires modeling subhalos well below the resolution limit of cosmological $N$-body simulations.
The properties of these unresolved subhalos cannot be directly determined from simulations.
Their contribution in Eq.~(\ref{eq:Btotal}) requires additional modeling~\cite{Sanchez-Conde:2013yxa,Bartels:2015uba}.
Here, we use the semi-analytical framework SASHIMI-C~\cite{Hiroshima:2018kfv,Ando:2019xlm}\footnote{\url{https://github.com/shinichiroando/sashimi-c}} to model the subhalo population.
For a host halo of mass $M$ at redshift $z$,
the subhalo mass $m$ in SASHIMI-C ranges from $10^{-6}M_{\odot}$ to the most massive subhalo that survives at $z$.
Its abundance and internal structure are then used to calculate the subhalo boost factor $B_{\rm sh}(M,z)$.
This treatment allows unresolved subhalos to be incorporated through semi-analytical modeling.

\begin{figure}[htb!]
	\centering
	\includegraphics[width=15cm, height=10cm]{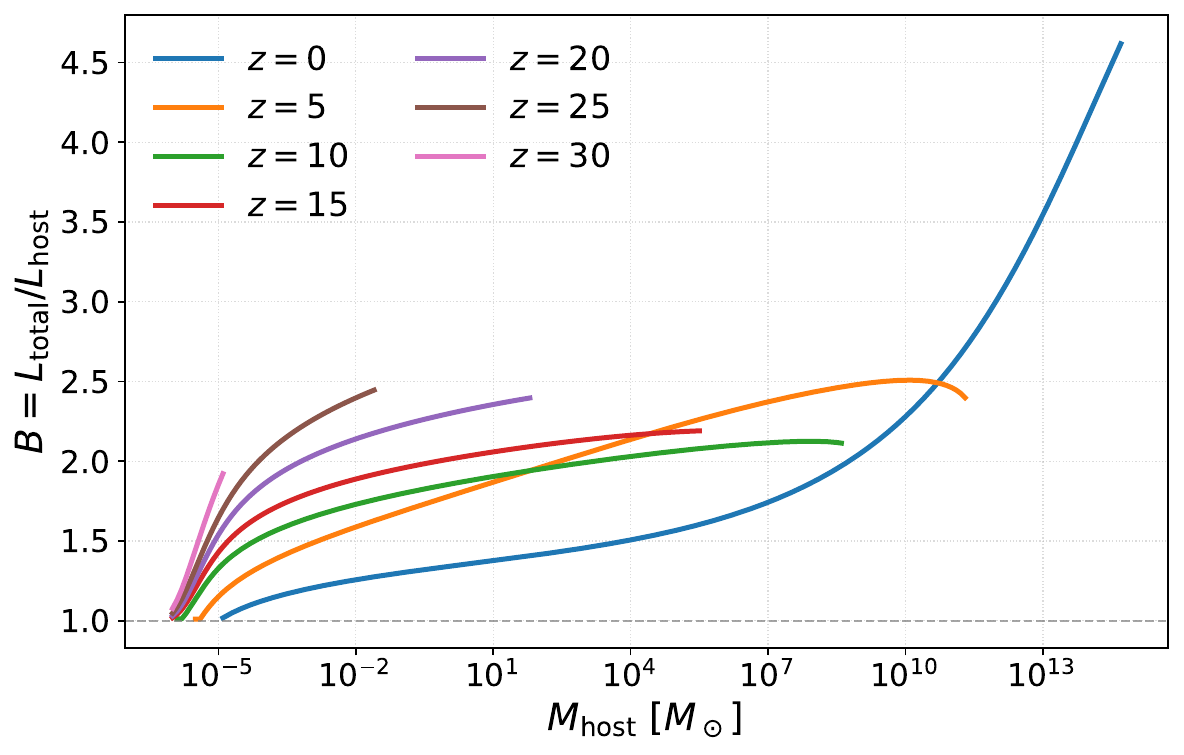}
	\centering
	 \caption{
The enhancement factor $B$ as a function of host-halo mass for selected redshifts, using SASHIMI-C.
  }
    \label{fig:sashimi_boost}
\end{figure}

Figure \ref{fig:sashimi_boost} shows the values of $B(M,z)$ as a function of host-halo mass and redshift.
$B$ remains of order unity over much of the relevant mass range, 
but increases toward sufficiently massive host halos.
As seen in Figure \ref{fig:sashimi_boost}, the range of host-halo masses covered by the SASHIMI-C calculation becomes narrower at higher redshift.

\subsection{Spatially inhomogeneous energy injection to the IGM}
\label{subsec:inhomogeneous_injection}

Following the halo-based inhomogeneous treatment of \cite{Sun:2025ksr},
the local injection spectrum induced by the DM annihilation in halos is
 \begin{align}
\frac{dN_i}{dE\,dV\,dt}(z,\mathbf{x})
&=\frac{dN_{\rm inj}}{dV\,dt}(z,\mathbf{x})\frac{dN_i}{dE}(z)\nonumber\\
&=\frac{1}{2m_\chi}\frac{dN_i}{dE}(z)\int dM\,\frac{dN}{dM}(M|z,\mathbf{x})L_{\rm total}(M,z),
\label{eq:inhomogeneous_injection}
\end{align}
which serves as a spatially varying source term in \texttt{DM21cm}.

The term $dN_{\rm inj}/dVdt$ in Eq.(\ref{eq:inhomogeneous_injection}) denotes the annihilation-event rate per unit physical volume and time,
with $dN_i/dE$ the spectrum of secondary photons ($i=\gamma$) or electrons and positrons ($i=e$) produced per annihilation event.
Since each annihilation of self-conjugate DM releases an energy $2m_\chi$,
the factor $1/(2m_\chi)$ converts annihilation luminosity into an event rate.
For the monochromatic channels considered here, 
$dN_i/dE=2\delta_{\rm D}(E-E_\star)$, where $\delta_{\rm D}$ is the Dirac delta, 
and $E$ is the photon (electron or positron kinetic) energy and $E_\star=m_\chi$ $(m_\chi-m_e)$ for  $\gamma\gamma$ ($e^+e^-$). 

The term $dN/dM(M|z,\mathbf{x})$ in Eq.(\ref{eq:inhomogeneous_injection}) is the local conditional halo mass function which gives the number density of halos per unit mass and physical volume at position $\mathbf{x}$ and redshift $z$.
This conditional mass function is determined by the local overdensity $\delta(\mathbf{x},z)$ supplied by \texttt{21cmFAST} \cite{Mesinger:2010ne,Park:2018ljd},
using the extended Press--Schechter (EPS) prescription \cite{Press:1973iz,Bond:1990iw} adopted by DM21cm \cite{Sun:2025ksr}.
Spatial variations in $\delta(\mathbf{x},z)$ produce the spatially inhomogeneous annihilation source.\footnote{Note, 
the mean annihilation rate is calculated by spatially averaging the local rates in \cite{Sun:2025ksr}.}

Finally, the halo mass integral in Eq.(\ref{eq:inhomogeneous_injection}) is from $M_{\min}=10^{-6}M_\odot$ to the grid-scale mass $M_{\rm grid}\simeq3\times10^{11}\,M_\odot$ that corresponds to the $2\,{\rm cMpc}$ comoving cell size adopted here.
For host halo masses outside the mass range covered by SASHIMI-C we set $B=1$.

\section{Impacts on the 21-cm Signal}
\label{sec:impact}

We now address how the subhalo-enhanced $s$-wave DM annihilation affects the 21-cm signal.

First, the spatially dependent injection spectra constructed in Eq.~(\ref{eq:inhomogeneous_injection}) provide the photon
and electron source terms for the subsequent propagation
and energy-deposition calculation in \texttt{DM21cm}.
Multiplying Eq.~(\ref{eq:inhomogeneous_injection}) by $\Delta t/\bar n_B(z)$, 
with $\bar n_B(z)$ the mean physical baryon number density and $\Delta t$ the proper-time interval, gives the injected spectrum during the evolution step \cite{Sun:2023acy}
\begin{equation}
\frac{dN_i^{\rm in}}{dE}(z,\mathbf{x})=\frac{\Delta t}{m_\chi\bar n_B(z)}\left[\int_{M_{\min}}^{M_{\rm grid}}dM \frac{dN}{dM}(M|z,\mathbf{x})L_{\rm total}(M,z)\right]\delta_{\rm D}(E-E_\star),
\label{eq:input_spectrum}
\end{equation}
where $dN_i/dE=2\delta_{\rm D}(E-E_\star)$ has been used.
Since $L_{\rm{total}}\sim \langle\sigma v\rangle/m_\chi$, 
Eq.(\ref{eq:input_spectrum}) suggests that $dN_i^{\rm in}/dE$ scales as $\sim \langle\sigma v\rangle/m_\chi^2$.
Then the evolution of the injected particles can be written as \cite{Sun:2023acy}
\begin{align}
\frac{dN_{\gamma}^{\rm out}}{dE}
&=T_{\gamma i}(\delta,x_{\rm HI}|z,\Delta z)\frac{dN_i^{\rm in}}{dE},
\label{eq:particle_propagation}\\
\begin{pmatrix}\Delta T_{k}\\\Delta x_e\\J_{\alpha}
\end{pmatrix}
&=D_{ci}(\delta,x_{\rm HI}|z,\Delta z)\frac{dN_i^{\rm in}}{dE},\qquad i=\gamma,e.
\label{eq:particle_deposition}
\end{align}
where $dN_{\gamma}^{\rm out}/dE$ is the photon spectrum per mean baryon after the corresponding propagation step.
The transfer function $T_{\gamma i}$ describes photon production and propagation, 
including surviving and scattered photons for $i=\gamma$,
which depends on the local matter overdensity $\delta$, neutral hydrogen fraction $x_{\rm HI}$,
redshift $z$, and evolution interval $\Delta z$.
The deposition transfer function $D_{ci}$ gives the  contributions to the gas kinetic temperature $T_k$, 
free-electron fraction $x_e$, and DM-induced Lyman-$\alpha$ intensity $J_\alpha$,
with the index $c$ denoting heating, ionization, and Lyman-$\alpha$ excitation.
Specifically, electron cooling is treated by the on-the-spot approximation \cite{Sun:2023acy,Liu:2019bbm}, 
whereas primary and secondary photons can propagate over cosmological distances before depositing their energy \cite{Evoli:2012zz}.

Second, these deposited contributions are incorporated into the standard thermal 
and ionization evolution in \texttt{21cmFAST}.
Additional ionization modifies $x_e$ and hence the neutral hydrogen fraction~\cite{Sun:2025ksr},
\begin{equation}
x_{\rm HI}=\max\!\left[0,x_{\rm HI}^{\rm filter}-x_e\right],
\label{eq:xHI}
\end{equation}
where $x_{\rm HI}^{\rm filter}$ is the neutral fraction obtained from the filter-based 
excursion-set treatment of UV-driven reionization~\cite{Park:2018ljd,Mesinger:2007pd}.
Moreover, the deposited heat modifies $T_{k}$, and the deposited Lyman-$\alpha$ contribution enters the Wouthuysen--Field coupling. 
All of these effects modify the hydrogen spin temperature \cite{Field:1959zz,Hirata:2005mz} via 
\begin{equation}
T_{\rm S}^{-1}=\frac{T_{\rm CMB}^{-1}+(x_{\rm c}+x_{\alpha})T_{k}^{-1}}{1+x_{\rm c}+x_{\alpha}},
\label{eq:Ts}
\end{equation}
where $T_{\rm CMB}$ is the CMB temperature, and $x_{\rm c}$ and
$x_{\alpha}$ are the collisional and Lyman-$\alpha$ coupling coefficients, respectively~\cite{Zygelman:2005gj,Furlanetto:2006su}.

Finally, the changes in $x_{\rm HI}$ and $T_{\rm S}$ are directly reflected 
in the differential 21-cm brightness temperature relative to the CMB~\cite{Furlanetto:2006jb,Pritchard:2011xb},
\begin{equation}
T_{21}=27\,x_{\rm HI}(1+\delta_{\rm b})\left(1-\frac{T_{\rm CMB}}{T_{\rm S}}\right)
\left(\frac{\Omega_{\rm b}h^2}{0.023}\right)
\left[\left(\frac{0.15}{\Omega_{\rm m}h^2}\right)\left(\frac{1+z}{10}\right)\right]^{1/2}
\left[\frac{H(z)}{H(z)+dv_r/dr}\right]{\rm mK},
\label{eq:T21}
\end{equation}
where $\delta_{\rm b}$ is the baryon overdensity,
$\Omega_{\rm b}$ and $\Omega_{\rm m}$ are the baryon and total matter density parameters, 
$h$ is defined by $H_0=100h\,{\rm km\,s^{-1}\,Mpc^{-1}}$, 
and $dv_r/dr$ is the comoving line-of-sight gradient of the peculiar velocity.

To quantify the subhalo effects on the 21-cm observables, we compare three scenarios:
(i) a baseline case with standard astrophysical evolution and no DM annihilation,
(ii) a host-only case including $s$-wave DM annihilation within smooth host halos,
and (iii) a host+subhalo case including $s$-wave DM annihilation with subhalo effects as described in Sec.~\ref{sec:swave_halo}.
Unless otherwise stated, we adopt a $2\,{\rm Mpc}$ comoving cell size,
consistent with the spatial resolution used for the local overdensity field in \texttt{DM21cm}.
All three cases use identical initial conditions and astrophysical and cosmological parameters.

\subsection{Thermal and ionization histories of IGM}

\begin{figure}[htb!]
    \centering
    \includegraphics[width=0.48\textwidth]{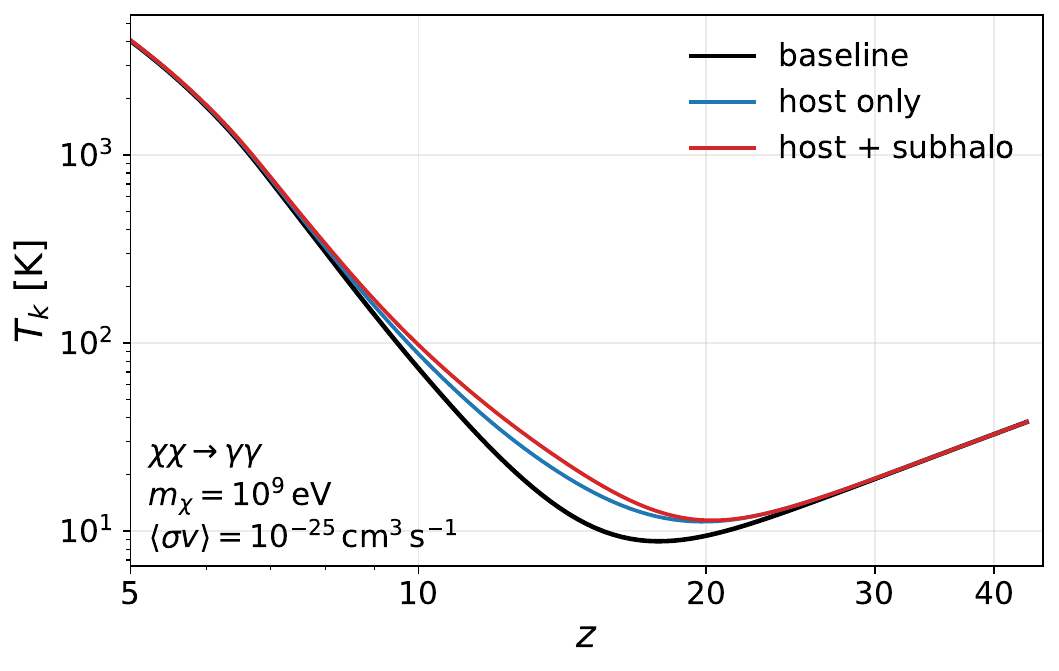}
    \hfill
    \includegraphics[width=0.48\textwidth]{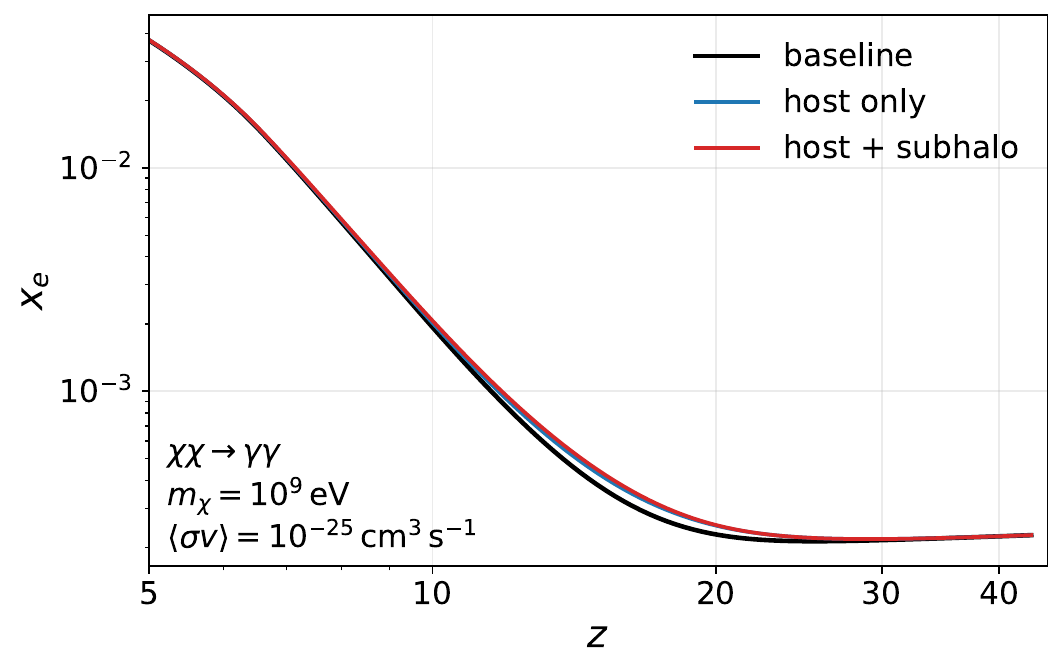}
    \caption{
    Evolution of the gas kinetic temperature $T_k$ (left) and free-electron fraction $x_e$ (right) 
    for the $\chi\chi\rightarrow\gamma\gamma$ annihilation channel with $m_\chi=10^9\,{\rm eV}$ 
    and $\langle\sigma v\rangle=10^{-25}\,{\rm cm^3\,s^{-1}}$. 
    The black, blue, and red curves correspond to the baseline, 
    host-only, and host+subhalo case, respectively.
    }
    \label{fig:thermal_phot}
\end{figure}

Figure \ref{fig:thermal_phot} shows the evolution of $T_k$ and $x_e$ for the $\chi\chi\rightarrow\gamma\gamma$ channel with the benchmark values of $m_\chi=10^9\,{\rm eV}$ and $\langle\sigma v\rangle=10^{-25}\,{\rm cm^3\,s^{-1}}$.
This figure shows that the contributions to $x_e$ and $T_k$ due to the DM annihilation become visible as the structure formation proceeds. 
In particular, the DM-annihilation-induced increase in the values of $T_k$ relative to its baseline values is noticeable at $z\sim 10-20$  in the host-only case, 
and even more significant in the host+subhalo case. 
By contrast, the increase in the value of $x_e$ is only mild both in the host-only and host+subhalo cases.

\begin{figure}[htb!]
    \centering
    \includegraphics[width=0.48\textwidth]{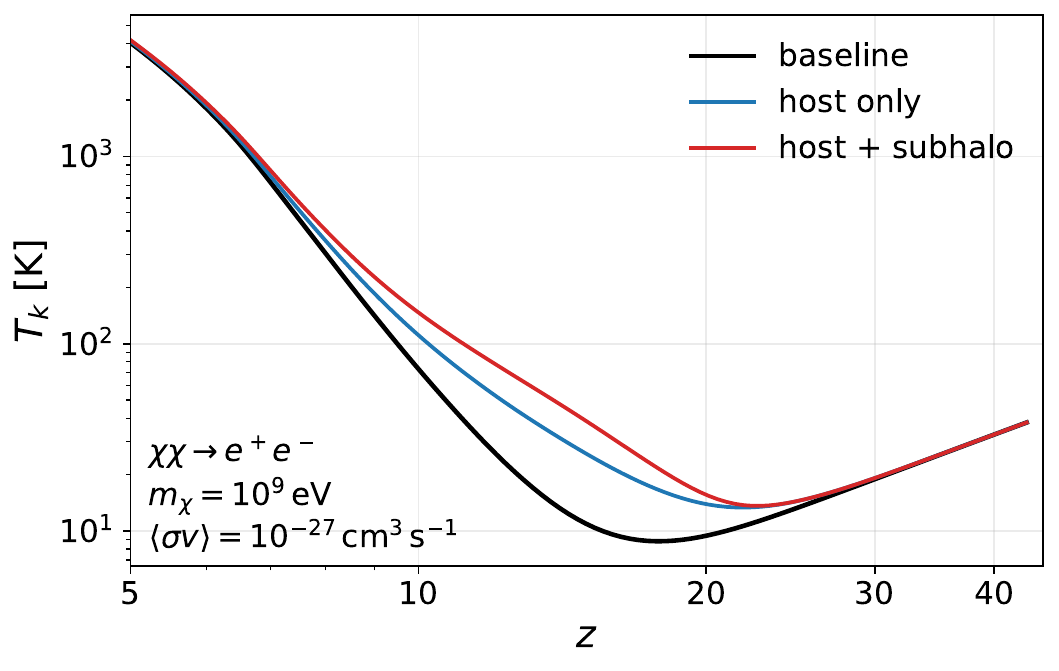}
    \hfill
    \includegraphics[width=0.48\textwidth]{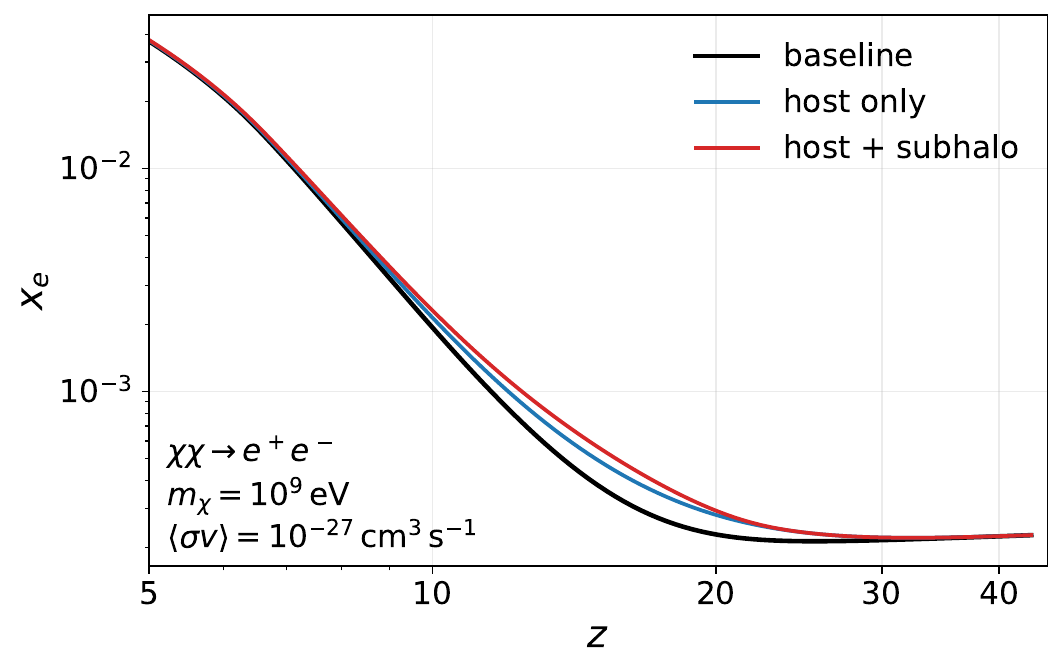}
    \caption{
    Same as Fig \ref{fig:thermal_phot}, but for the $\chi\chi\rightarrow e^+e^-$ annihilation channel 
    with $m_\chi=10^9\,{\rm eV}$ and $\langle\sigma v\rangle=10^{-27}\,{\rm cm^3\,s^{-1}}$.}
    \label{fig:thermal_elec}
\end{figure}

Similar to Figure \ref{fig:thermal_phot}, we show in Figure \ref{fig:thermal_elec} the evolution of $T_k$ and $x_e$ for the $\chi\chi\rightarrow e^+e^-$ annihilation channel with the benchmark values of $m_\chi=10^9\,{\rm eV}$ and $\langle\sigma v\rangle=10^{-27}\,{\rm cm^3\,s^{-1}}$.
A comparison between Figures \ref{fig:thermal_phot} and \ref{fig:thermal_elec} illustrates (i) the DM annihilation results in increases of $T_k$ and $x_e$ both in the host-only and host+subhalo cases; and (ii) the efficiency of energy deposition is higher through the $e^{+}e^{-}$ channel than the $\gamma\gamma$ channel, as electrons transfer energy to CMB photons through inverse Compton scattering, producing a secondary spectrum \cite{Liu:2019bbm,Sun:2023acy} of photons with shorter absorption lengths than directly injected photons.

\subsection{21-cm brightness temperature and lightcones}

\begin{figure}[htb!]
    \centering
    \includegraphics[width=0.48\textwidth]{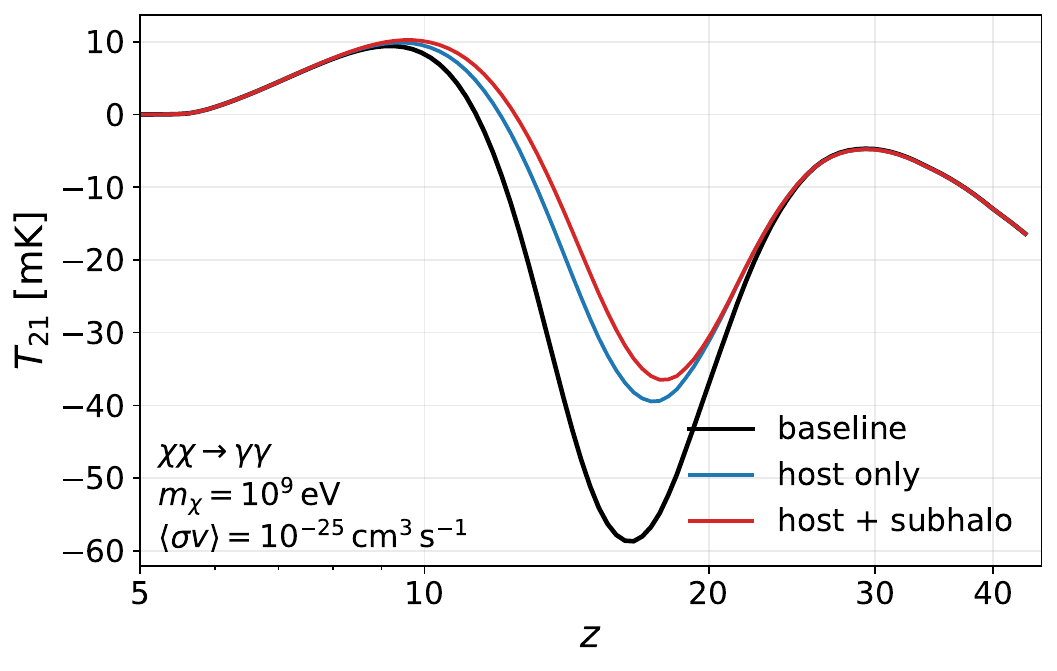}
    \hfill
    \includegraphics[width=0.48\textwidth]{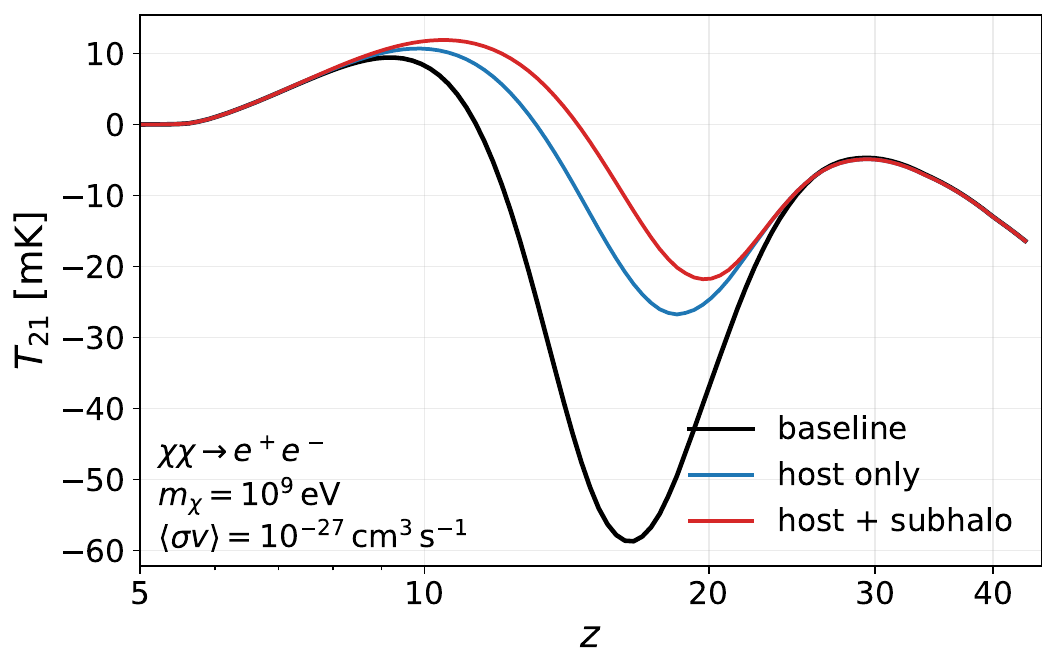}
    \caption{
    Sky-averaged 21-cm brightness temperature for $\chi\chi\rightarrow\gamma\gamma$ 
    with $m_\chi=10^9\,{\rm eV}$ and $\langle\sigma v\rangle=10^{-25}\,{\rm cm^3\,s^{-1}}$ (left), 
    and for $\chi\chi\rightarrow e^+e^-$ with $m_\chi=10^9\,{\rm eV}$ 
    and $\langle\sigma v\rangle=10^{-27}\,{\rm cm^3\,s^{-1}}$ (right). 
    The black, blue, and red curves correspond to the baseline, host-only, and host+subhalo case, respectively.
    }
    \label{fig:T21}
\end{figure}

Figure \ref{fig:T21} shows the sky-averaged 21-cm brightness temperature $T_{21}$ for the two annihilation channels discussed above. 
Either in the $\chi\chi\rightarrow\gamma\gamma$ (left) or  $\chi\chi\rightarrow e^+e^-$ (right) channel,  
the values of $T_{21}$ during the cosmic dawn are uplifted both in the host-only and host+subhalo case,
as the DM annihilation induced heating of the IGM can drive the spin temperature upward, and therefore reduce the absorption depth of $T_{21}$.
Compared to the host-only case, the increase in the values of  $T_{21}$ in either of the two annihilation channels is more obvious in the host+subhalo case, 
verifying that the subhalo effects on the 21-cm observables cannot be neglected.

\begin{figure}[p!]
    \centering
    \includegraphics[width=0.85\textwidth]{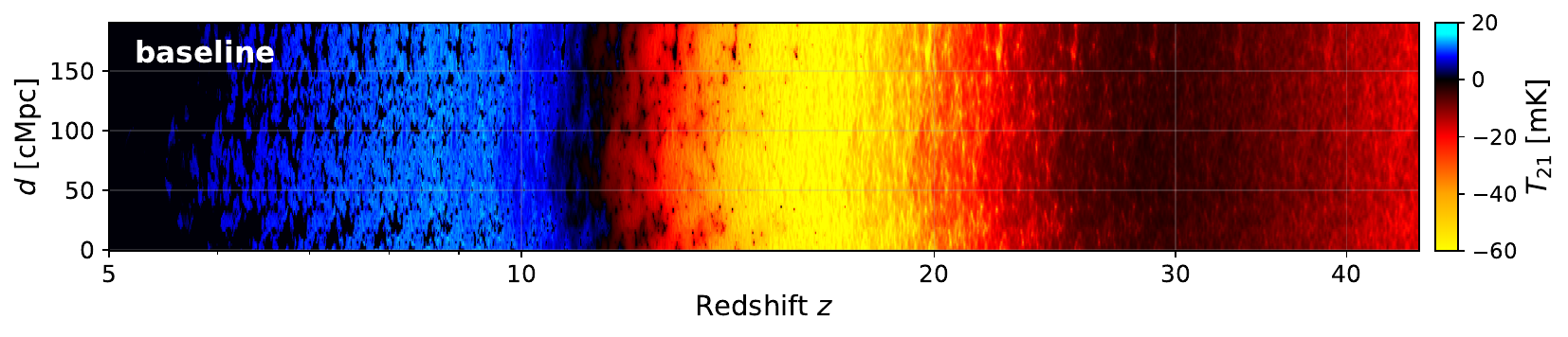}
    \includegraphics[width=0.85\textwidth]{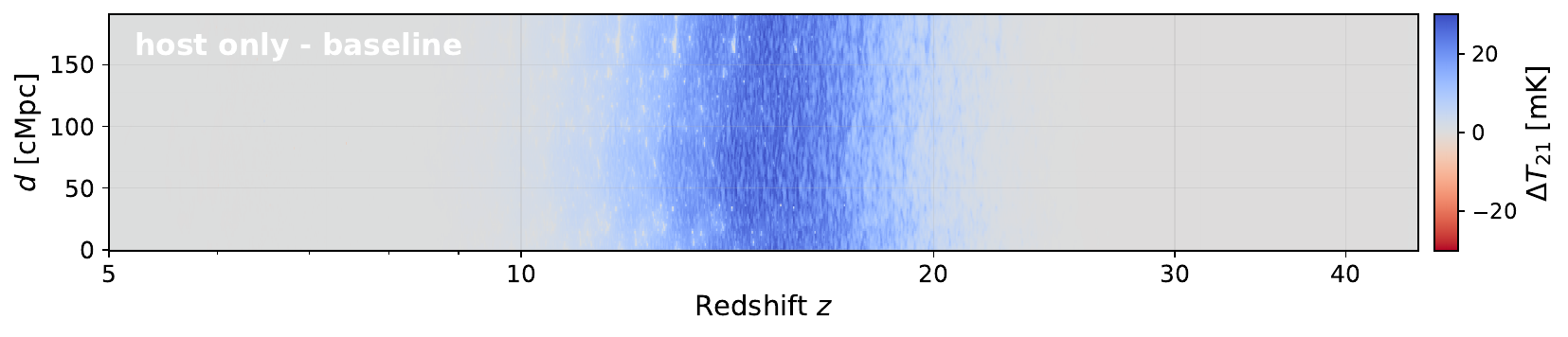}
    \includegraphics[width=0.85\textwidth]{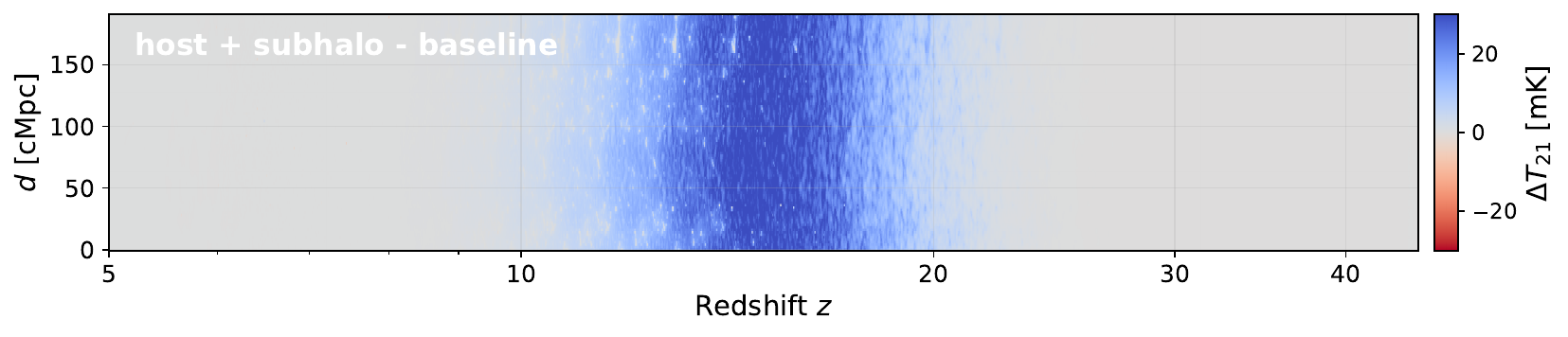}
    \caption{
    21-cm brightness-temperature lightcones for $\chi\chi\rightarrow\gamma\gamma$ 
    with $m_\chi=10^9\,{\rm eV}$ and $\langle\sigma v\rangle=10^{-25}\,{\rm cm^3\,s^{-1}}$. 
    From top to bottom, the panels show the baseline lightcones, the residuals relative to the baseline lightcones in the host-only and host+subhalo case, respectively.
    }
    \label{fig:lightcone_phot}
\end{figure}

\begin{figure}[p!]
    \centering
    \includegraphics[width=0.85\textwidth]{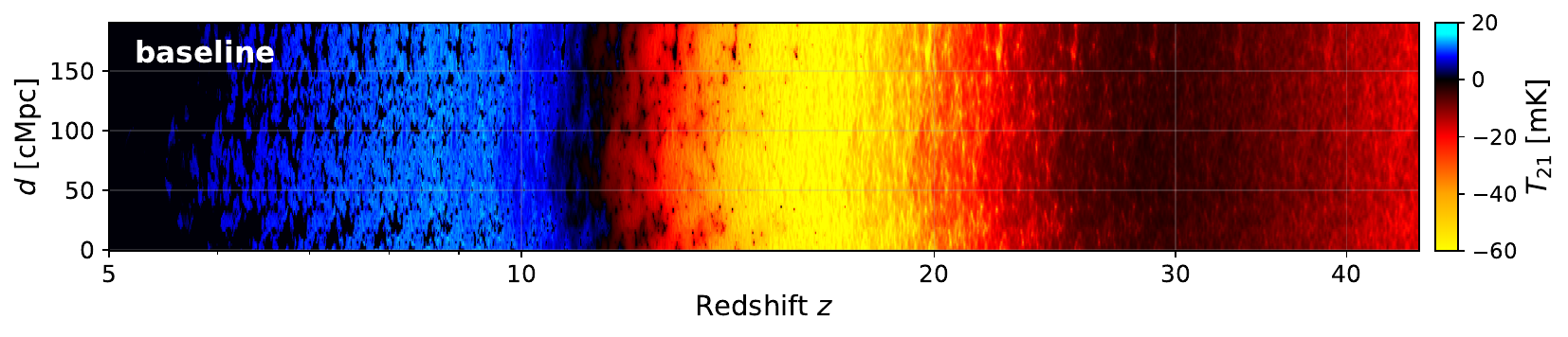}
    \includegraphics[width=0.85\textwidth]{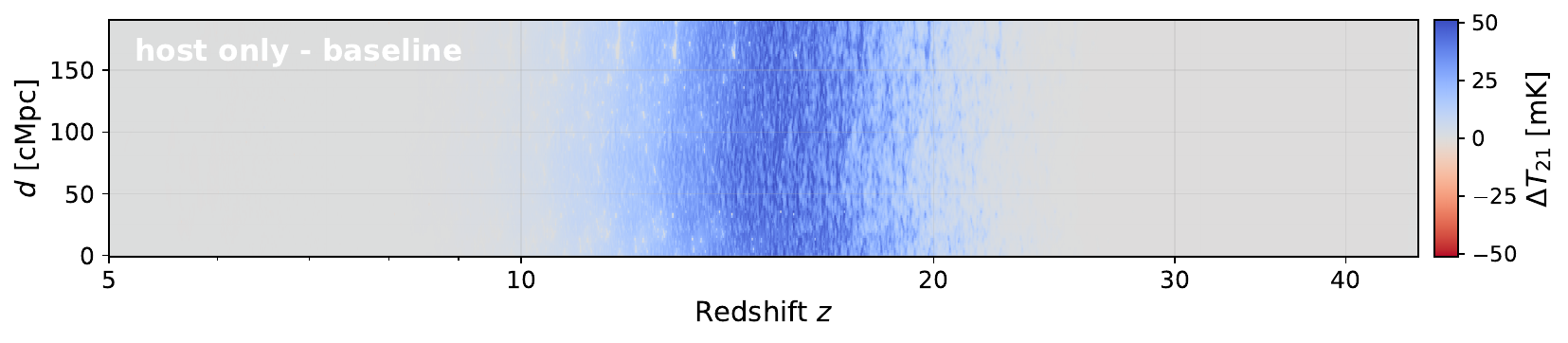}
    \includegraphics[width=0.85\textwidth]{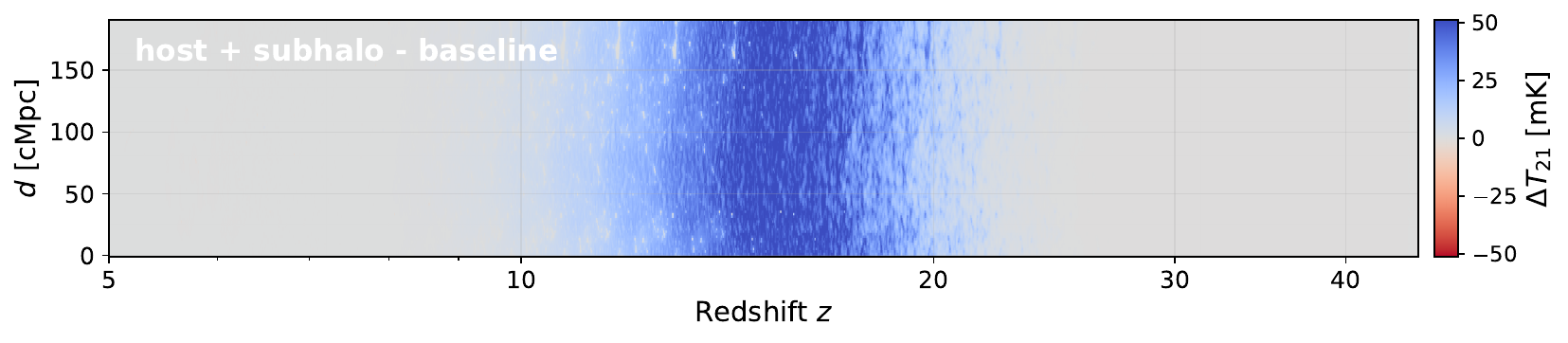}
    \caption{
    Same as Fig.~\ref{fig:lightcone_phot}, but for $\chi\chi\rightarrow e^+e^-$ with $m_\chi=10^9\,{\rm eV}$ 
    and $\langle\sigma v\rangle=10^{-27}\,{\rm cm^3\,s^{-1}}$.
    }
    \label{fig:lightcone_elec}
\end{figure}

Figures \ref{fig:lightcone_phot} and \ref{fig:lightcone_elec} show the spatial evolution 
of the 21-cm brightness temperature for the two annihilation channels. 
Therein the baseline lightcone corresponds to the standard astrophysical evolution without energy injection from the DM annihilation. 
In each figure the residuals defined as $T_{21}^{\rm host}-T_{21}^{\rm baseline}$ (middle) and $T_{21}^{\rm host+subhalo}-T_{21}^{\rm baseline}$ (bottom) 
are shown for comparison.
The results are consistent with those in Figure \ref{fig:T21}, i.e., 
new energy deposition due to the DM annihilation raises the gas temperature and reduces the depth of the 21-cm absorption signal.
Besides, the residual lightcones also show that the effects of DM annihilation are spatially dependent, 
as expected from the inhomogeneous annihilation source constructed from the local halo abundance.
These spatially dependent modifications affect the 21-cm fluctuations and are relevant to the power-spectrum sensitivity forecasts discussed below.

The electron channel in Figure \ref{fig:lightcone_elec} exhibits a larger residual amplitude than the photon channel in Figure \ref{fig:lightcone_phot}, 
which is consistent with the larger separation of the sky-averaged curves as shown by the two plots in Figure \ref{fig:T21}.

\section{Projected constraints}
\label{sec:constraints}

In this section, we derive the projected 21-cm constraints on the $s$-wave DM annihilation cross section\DIFaddbegin \DIFadd{.
}\DIFaddend We use the Fisher-matrix framework of \texttt{21cmfish} \cite{Mason:2022obt}.
 Regarding the astrophysical background, we adopt the fiducial ``best-guess'' model in the \texttt{21cmFAST} forecasts \cite{Munoz:2021psm},
where the astrophysical nuisance parameters and their fiducial values are summarized in Table \ref{tab:astro_params}.
We use the moderate foreground prescription implemented in \texttt{21cmSense}~\cite{Murray:2024the}. 
For the HERA sensitivity, we follow the forecast configuration of \cite{Mason:2022obt,Sun:2025ksr},
assuming 331 antennas and a total observing time of $1080$ hours.

\begin{table}
\centering
\begin{tabular}{lcccc}
\hline\hline
PopII parameters
& $f_{\star,10}^{\rm II}$
& $\alpha_\star^{\rm II}$
& $f_{\rm esc,10}^{\rm II}$
& $L_X^{\rm II}$ \\
Fiducial value
& $-1.25$
& $0.5$
& $-1.35$
& $40.5$ \\
\hline
PopIII parameters
& $f_{\star,7}^{\rm III}$
& $\alpha_\star^{\rm III}$
& $f_{\rm esc,7}^{\rm III}$
& $L_X^{\rm III}$ \\
Fiducial value
& $-2.5$
& $0.0$
& $-1.35$
& $40.5$ \\
\hline
Shared parameters
& $t_\star$
& $\alpha_{\rm esc}$
& $E_0$
& $A_{\rm LW}$ \\
Fiducial value
& $0.5$
& $-0.3$
& $500$
& $2.0$ \\
\hline\hline
\end{tabular}
\caption{Astrophysical nuisance parameters and their fiducial values
adopted in the Fisher forecast following Ref.\cite{Munoz:2021psm}.}
\label{tab:astro_params}
\end{table}

We take the velocity-averaged $s$-wave DM annihilation cross section,
$A\equiv\langle\sigma v\rangle$, as an additional Fisher parameter.
Combining $A$ with the astrophysical nuisance parameters listed in Table \ref{tab:astro_params} forms the full Fisher parameter set.
In terms of the Fisher parameter set, we define the Fisher matrix as \cite{Sun:2025ksr}
\begin{equation}
F_{ij}=\sum_{k,z}
\frac{\partial_i\mathcal P(k,z)\,\partial_j\mathcal P(k,z)}{\sigma_{\rm obs}^2(k,z)+\sigma_{\rm sim}^2(k,z)+[0.2\mathcal P_0(k,z)]^2},
\label{eq:Fisher}
\end{equation}
where the 21-cm power spectrum $\mathcal P(k,z)=k^3P_{21}(k,z)/(2\pi^2)$ with $P_{21}$ the brightness-temperature power spectrum,
$\partial_i=\partial/\partial\theta_i$ with $\theta_i$ denoting the Fisher parameters, 
 $\sigma_{\rm obs}$ and $\sigma_{\rm sim}$ describe the observational and finite-volume simulation uncertainties respectively, 
the last term in the denominator represents a $20\%$ modeling uncertainty \cite{Mason:2022obt,Sun:2025ksr} with $\mathcal P_0$ the fiducial power spectrum,
and finally the sum runs over independent wavenumber and redshift bins. 
Specifically, the derivative of $\mathcal P(k,z)$ over  $A$ is evaluated 
using a second-order forward finite difference at $A=0$, $\Delta A$, and $2\Delta A$.

After marginalizing over the astrophysical nuisance parameters,
the projected $95\%$ upper limit is
\begin{equation}
\langle\sigma v\rangle_{95}
=1.65\sqrt{\left(F^{-1}\right)_{AA}},
\label{eq:sv_limit}
\end{equation}
where the factor $1.65$ corresponds to the one-sided $95\%$ confidence level (CL) for a Gaussian likelihood.

\begin{figure*}
    \centering
    \includegraphics[width=0.48\textwidth]{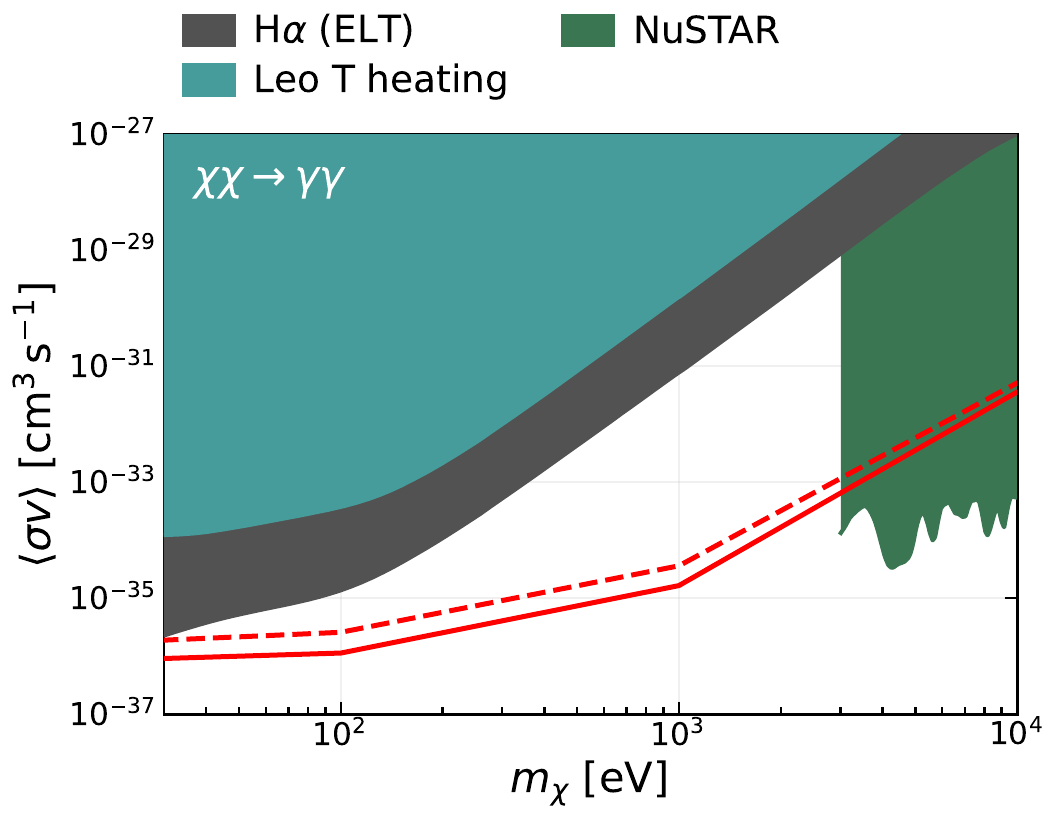}
    \hfill
    \includegraphics[width=0.48\textwidth]{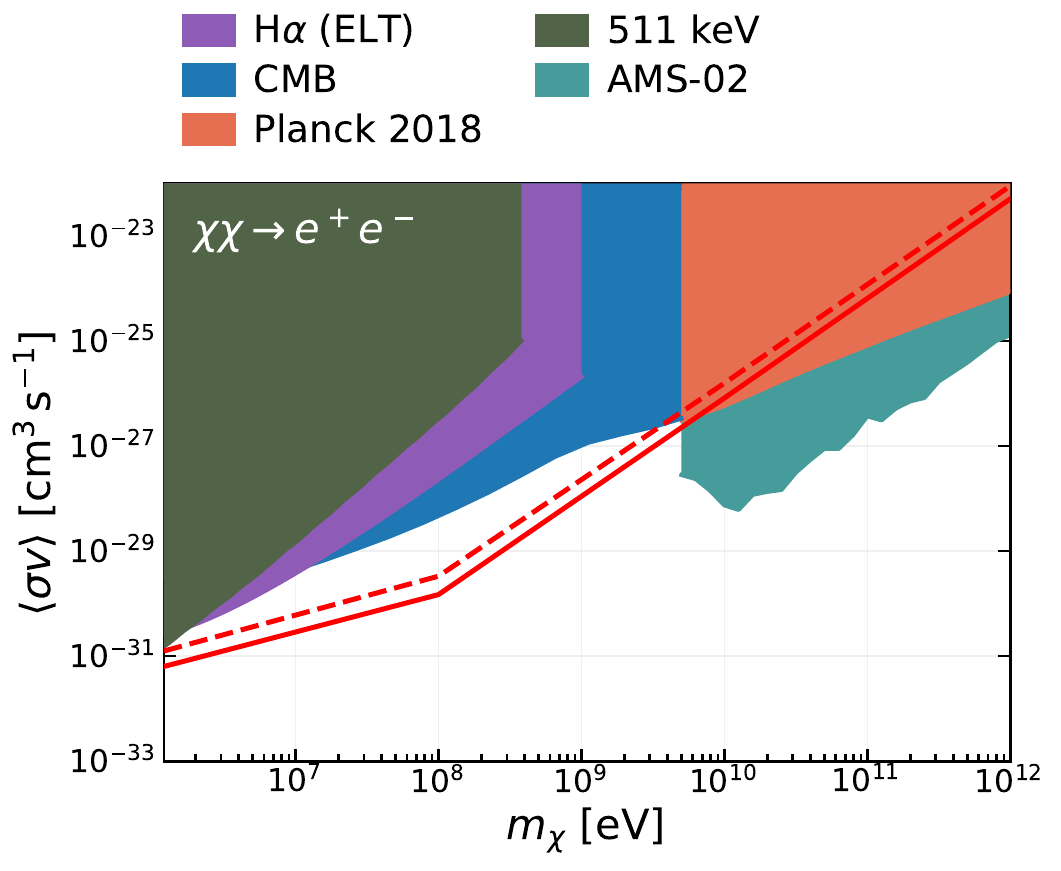}
    \caption{
 Projected $95\%$ upper limits on the velocity-averaged
$s$-wave DM annihilation cross section as a function of the DM mass.
The left panel shows the $\chi\chi\rightarrow\gamma\gamma$ channel
compared to the Leo~T bound \cite{Wadekar:2021qae} as corrected by Ref.\cite{Leane:2025ohu},
the projected H$\alpha$ sensitivity of the extremely large telescope (ELT) \cite{Leane:2025ohu}, 
and the NuSTAR bound \cite{Zakharov:2025coj}.
The right panel shows the $\chi\chi\rightarrow e^+e^-$ channel
compared to the limits of H$\alpha$ (ELT) \cite{Leane:2025ohu}, CMB~\cite{Slatyer:2015jla},
Planck 2018~\cite{Planck:2018vyg}, Galactic $511\,{\rm keV}$ line \cite{DelaTorreLuque:2023cef},
and AMS-02 \cite{John:2021ugy}.
Here, the dashed and solid red curves refer to the HERA forecasts in the host-only and host+subhalo case, respectively.}
\label{fig:swave_constraints}
\end{figure*}

Figure \ref{fig:swave_constraints} shows the projected $95\%$ CL upper limits on $\langle\sigma v\rangle$ of $s$-wave DM annihilation
$\chi\chi\rightarrow\gamma\gamma$ (left) and $\chi\chi\rightarrow e^+e^-$ (right),
which are compared to the existing astrophysical and cosmological bounds.
The dashed and solid red curves refer to the host-only and host+subhalo cases, respectively.

Regarding the $\gamma\gamma$ channel as shown in the left panel of Figure \ref{fig:swave_constraints},
the HERA limits in the mass range of $m_\chi\sim0.1$--$3\,{\rm keV}$ are the strongest compared to the Leo~T bound \cite{Wadekar:2021qae} as corrected by Ref.\cite{Leane:2025ohu},
the H$\alpha$ sensitivity (ELT) \cite{Leane:2025ohu} and the NuSTAR bound \cite{Zakharov:2025coj}.
In the mass range of $m_{\chi}\sim3$--$10\,{\rm keV}$, the NuSTAR bound is however stronger than the HERA limits.
For larger DM masses, the HERA limits become much weaker than the existing bounds, 
because the DM annihilation injects less energy for larger DM masses given a fixed value of $\langle\sigma v\rangle$.
Therefore, these large DM mass regions have been neglected for simplicity.

 For the $e^+e^-$ channel as shown in the right panel of Figure \ref{fig:swave_constraints},
 the HERA limits are rather sensitive to the DM mass.
In the mass range of $m_\chi\sim 1$\rm{MeV}--$3\,{\rm GeV}$, 
the HERA limits are the strongest compared to the existing limits of H$\alpha$ (ELT) \cite{Leane:2025ohu}, CMB~\cite{Slatyer:2015jla},
Planck 2018~\cite{Planck:2018vyg}, Galactic $511\,{\rm keV}$ line \cite{DelaTorreLuque:2023cef} and AMS-02 \cite{John:2021ugy}.
In the mass range of $m>10$ GeV, 
 the HERA limits  are weaker than the AMS-02 bound \cite{John:2021ugy}, 
 but still comparable with the Planck 2018 limit~\cite{Planck:2018vyg}.

In each of the two channels as shown in Figure \ref{fig:swave_constraints} 
the HERA limits on the $s$-wave DM annihilation cross section are strengthened
by a factor of  about two in the host+subhalo case relative to the host-only case, 
as a result of the enhancement on the annihilation luminosity due to the halo substructure.

\section{Conclusions}
\label{sec:conclusion}

In this work, we have investigated the impacts of DM substructure on the 21-cm observables by
extending the halo-based treatment of $p$-wave DM annihilation in \texttt{DM21cm} to the velocity-independent $s$-wave DM annihilation.
To this end, we have adopted SASHIMI-C to handle the subhalo population,
which enables the spatially inhomogeneous DM annihilation to include the subhalo effects.
Based on these results, we have studied the subhalo effects on the thermal and ionization histories of the IGM and the 21-cm brightness temperature and power spectra.

For the two explicit annihilation channels $\chi\chi\rightarrow\gamma\gamma$ and $\chi\chi\rightarrow e^+e^-$ considered in this work,
we have shown that the projected HERA limits on the $s$-wave DM annihilation cross section have been strengthened by a factor of about two in the host+subhalo case relative to the host-only case.
Explicitly, the HERA limits for the $\gamma\gamma$ channel are the strongest in the mass range of $m_\chi\sim0.1$--$3\,{\rm keV}$ compared to 
the existing limits of Leo~T,  H$\alpha$ (ELT) and NuSTAR;
whereas the HERA limits for the $e^+e^-$ channel are the strongest in the mass range of $m_\chi\sim 1$ MeV --$3$ GeV compared to 
the existing limits of H$\alpha$ (ELT), CMB, Planck 2018, Galactic $511\,{\rm keV}$ line and AMS-02.

There are a few interesting points left for future study.
First, the 21-cm constraints on the s-wave DM annihilation cross section, which are DM model independent, 
can be applied to explicit models such as DM via the vector boson portal and a scalar DM through the Higgs portal. 
Moreover, the future 21-cm data can be used to constrain the density and structure of subhalos.

\section*{Acknowledgements}
The authors acknowledge the use of \texttt{DM21cm}~\cite{Sun:2023acy}, \texttt{DarkHistory}~\cite{Liu:2019bbm},
\texttt{21cmFAST}~\cite{Mesinger:2010ne}, \texttt{SASHIMI-C} \cite{Hiroshima:2018kfv,Ando:2019xlm}, 
\texttt{hmf} \cite{Murray:2013qza}, \texttt{halomod} \cite{Murray:2020dcd},
\texttt{21cmfish}~\cite{Mason:2022obt}, and \texttt{21cmSense}~\cite{Murray:2024the}.
The codes are available at the website \texttt{https://github.com/hunt4dm/dm21cm-sashimi-swave}.


\end{document}